\documentstyle[preprint,eqsecnum,aps,prb]{revtex}

\begin{document}
\title{Correlation between structure and properties in multiferroic La$_{0.7}$Ca$%
_{0.3}$MnO$_3$/BaTiO$_3$ superlattices. }
\author{M.P.\ Singh$^1$, W. Prellier$^{1,}$\thanks{%
prellier@ensicaen.fr}, L.\ Mechin$^2$, Ch.\ Simon$^1$ and B.\ Raveau$^1$.}
\address{$^1$Laboratoire CRISMAT, CNRS\ UMR 6508, ENSICAEN,\\
6 Bd du Mar\'{e}chal Juin, F-14050 Caen Cedex, FRANCE.\\
$^2$Laboratoire GREYC, CNRS\ UMR 6072, ENSICAEN and University of Caen, 6 Bd%
\\
du Mar\'{e}chal Juin, F-14050 Caen Cedex, FRANCE.}
\date{\today}
\maketitle

\begin{abstract}
Superlattices composed of ferromagnetics, namely La$_{0.7}$Ca$_{0.3}$MnO$_3$
(LCMO), and ferroelectrics, namely, BaTiO$_3$(BTO) were grown on SrTiO$_3$
at 720$^o$C by pulsed laser deposition process. While the out-of-plane
lattice parameters of the superlattices, as extracted from the X-ray
diffraction studies, were found to be dependent on the BTO layer thickness,
the in-plane lattice parameter is almost constant. The evolution of the
strains, their nature, and their distribution in the samples, were examined
by the conventional sin$^2\psi $ method. The effects of structural variation
on the physical properties, as well as the possible role of the strain on
inducing the multiferroism in the superlattices, have also been discussed.
\end{abstract}

\newpage

\section{Introduction}

Superlattices, which are composed of thin layers of two or more different
structural counterparts that are stacked in a well-defined sequence, may
exhibit some remarkable properties that do not exist in either of their
parent compounds. For example, a (LaFeO$_3$)/(LaCrO$_3$) superlattice
stacked on (111)-SrTiO$_3$ exhibits a ferromagnetic behavior, whereas each
parent material is antiferromagnetic.\cite{Kueda} Similarly, perovskite
based superlattices have shown new or enhanced properties such as high
temperature superconductivity, ferroelectricity, {\it etc.}\cite
{Norton,Gong,Yin,Ultrathin,Lu,mhjo,padhan,Liang} In recent years, various
efforts have been made to synthesize and study the structure and properties
of various superlattices based on perovskite oxides. Hence, the
superlattices offer a novel approach to create a new type of materials or
tailor the existing one for their suitability in real applications.
Generally, the properties of the superlattices depend on the various
factors, such as type of layer materials, their thickness, morphology,
interfacial structure between them, substrate, etc. Moreover, the lattice
mismatch among the individual type of layer materials employed for creating
the superlattices, and between the substrate and the individual type of
layers, play an important role in determining the interfacial structure
which in turn governs their electronic and magnetic properties.\cite
{Ultrathin}

Materials in which ferromagnetism and ferroelectricity coexist \cite{Schmid}
and eventually, possess multiferroic coupling, {\it i.e.,} magnetic domains
can be tuned by the application of an external electric field (and likewise
electric domains are switched by magnetic field) are known as multiferroics.
Thus, these materials offer an additional degree of freedom in designing the
various novel devices, {\it e.g.,} transducers, actuators, and storage
devices, which are unachievable separately in either ferroelectric or
magnetic materials.\cite{Hill} Hitherto, very few materials exist in nature
or synthesized in laboratory which exhibit multiferroism.\cite
{Sharan,wang,Lorenz,Sharma,hur,kimura,kimuraII}\ Why and under what
circumstances a large multiferroic coupling should come about is an open
question. However, this problem has been proven difficult to tackle owing to
the lack of materials which possess large multiferroic coupling. The
scarcity of multiferroics with large multiferroic coupling at moderate
conditions is also one of the big hurdle in the realization of multiferroic
devices. Thus, it is very important to design novel multiferroics with
essential properties.

To synthesize the artificial structures which exhibit multiferroism, various
approaches have been made. One of the most common approach, which can be
used for synthesizing the artificial multiferroics is doping of the magnetic
impurities in ferroelectric host. Other alternative direction which have
been adopted is to synthesize the composites by mixing the ferroelectrics
and ferromagnetic in the form of either bulk or thin films.\cite{zheng}
Superlattices approach that have been adopted for synthesizing the new
materials, can also be employed for designing multiferroics. In our previous
works, we have demonstrated that superlattices composed of ferromagnetic and
ferroelectric layers posses extraordinary magnetoelectrical properties and
such results have been understood based on the possible multiferroic
coupling in these structures.\cite{murga,murgaII,murgaIII,mpsingh}
Furthermore, we have also shown that their magnetoelectric properties are
depending on the nature layers, i.e. ferroelectric or paraelectric. For
example, a drastic enhancement in the magnetoelectric properties of Pr$%
_{0.85}$Ca$_{0.15}$MnO$_3$/Ba$_{0.6}$Sr$_{0.4}$TiO$_3$\ superlattices is
observed, whereas it was absent in Pr$_{0.85}$Ca$_{0.15}$MnO$_3$/SrTiO$_3$\
superlattices\cite{murgaIII} confirming the importance of the nature of the
layers for the multiferroic properties. Very recently, we have also shown
the presence of magnetocapacitance effects in La$_{0.7}$Ca$_{0.3}$MnO$_3$%
/BaTiO$_3$\ superlattices, which demonstrate that these superlattices are
behaving as multiferroics.\cite{mpsingh} Furthermore, La$_{0.7}$Ca$_{0.3}$MnO%
$_3$/BaTiO$_3$ superlattices also have exhibited an enhancement in their
magnetization with the progressive increase in the ferroelectric layer
thickness,\cite{mpsingh} indicating that such superlattices can exhibit
multiferroism under suitable conditions.

As discussed above, various recent studies on the superlattices shown that
their properties depend strongly on their structures, {\it e.g.,}
morphology, thickness, compositions, strains {\it etc}. Thus, it is expected
that the multiferroism in the superlattices will be dependent on their
structure, which in turn will govern their properties and consequently, the
performance of the devices based upon them. Therefore, we have investigated
the structure of superlattices composed of La$_{0.7}$Ca$_{0.3}$MnO$_3$
(LCMO) and BaTiO$_3$ (BTO) in order to study their structure. The evolution
of the strains as a function of the BTO layer thickness, and the structural
coherency have been analyzed by utilizing asymmetrical X-ray reflections and
the sin$^2\psi $ method. A correlation between structure and properties has
been established and our results are reported in this article.

\section{Experimental details}

The deposition of the La$_{0.7}$Ca$_{0.3}$MnO$_3$/BaTiO$_3$ (LCMO/BTO)
superlattices on (001)-oriented SrTiO$_3$ (STO) were carried out at 720$^o$C
in a flowing 100 mTorr O$_2$ ambient by a multitarget pulsed laser
deposition technique. Superlattices composed of individual BTO layer
thickness varying from 1 to 25 unit cells (u.c.) and the 5 u.c. LCMO with a
total periodicity of 25 were realized. 5\ u.c. of LCMO\ were chosen because
thin layers of LCMO\ behave as a ferromagnetic insulator, which is an
important factor in designing the artificial multiferroics.\cite{Blamire}

The epitaxial nature and the structural characterization of the films were
performed using the Seifert 3000P and Phillips X'Pert X-ray diffractometers
(Cu K$\alpha $ , $\lambda $= 0.15406 nm). The $\theta -2\theta ,\phi ,$ and
rocking curve ( $\omega $) -scans were recorded on the samples.
Morphological study of the films was carried out by atomic force microscopy.
Magnetization (M) of the films was measured as a function of temperature (T)
and magnetic field (H) using a superconducting quantum interference device
magnetometer (SQUID). $DC$ electrical properties of films were measured in
four-point probe configuration.

\section{Results and discussion}

\subsection{Structural studies}

Both LCMO and BTO possess the perovskite structure and their structures are
well studied and documented.\cite{jcpds} LCMO exhibits a cubic symmetry and
its bulk lattice parameter is 0.386 nm. Thus, the lattice mismatch between
the SrTiO$_3$ (STO) (a = 0.3905 nm) and LCMO is about $-1.17\%$. In
contrast, BTO exhibits several crystallographic polymorphs. The most stable
polymorphs are tetragonal (a = 0.39926 nm and c= 0.40309 nm) and cubic (a =
0.4006 nm). The lattice mismatch between BTO and STO is close to $+2.2\%$.

To investigate the crystallinity and epitaxial nature of the films, X-ray
diffraction (XRD) studies were carried out on various samples. Fig 1 shows $%
\theta -2\theta $ XRD patterns of the films around the (002) reflection with
varying BTO layer thickness. The presence of higher order satellite peaks
(denoted by the number $i$ which corresponds to the $i^{th}$ satellite peak)
adjacent to the main peak (denoted by $i$ = 0) indicates the formation of a
new structure having a periodic chemical modulation ($\Lambda $) of the
constituents and a coherently grown film, {\it i.e.,} with the single value
of in-plane lattice parameter throughout the entire thickness. The
out-of-plane lattice parameter of the superlattice ($\Lambda $), calculated
from $\theta -2\theta $ XRD patterns (Fig. 1b) is in good agreement with
theoretical values. To extract the information about the coherency at
interfaces, we have carried out a quantitative refinement of the
superlattice structure using the DIFFAX program.\cite{see} The experimental
and simulated diffraction profiles of the (5/10) superlattice structure are
shown in Fig 2a. The simulated profile is in close agreement with the
observed XRD pattern revealing a coherently grown structure.

Furthermore, to examine the in-plane coherence, $\Phi $-scan was recorded
around the $103$-reflection of the superlattices. A typical pole figures (%
{\it i.e}. a series of $\Phi -$scans plotted in three dimensions) for (5/10)
superlattice is shown in Fig 2b. Four peaks are clearly observed at $90{%
{}^{\circ }}$ from each other, indicating a four-fold symmetry as expected
for the perovskite structures- LCMO\ and BTO. Similar scan recorded on the
substrate confirms that the superlattice grown epitaxially ''cube-on-cube''.
Thus, from the observed XRD pattern it is evident that films have
well-defined interfaces. To get more in-depth information about the strain
and as well as coherence in the films, rocking curves around the
002-reflection of the film were recorded. A typical rocking curve for a 5/10
superlattice is shown in the inset of Fig 3a. The observed
full-width-at-half-maximum (FWHM) of the rocking curve recorded around the
fundamental ($002$) diffraction peak is close to the instrumental broadening
($<0.3{^{\circ }}$), indicating a good crystallinity and a good coherency.
The FWHM as a function of BTO u.c. is plotted in Fig 3a. It shows that the
FWHM of the film is varying with the BTO thickness and attains maximum value
(\symbol{126}0.2$^{\circ }$) for the (5/15) superlattice. The FWHM of the
XRD peak in thin films usually comes from various factors, {\it e.g.}
crystallite size, strain, defects, substrates and film/substrate interface
etc. Moreover, in the case of epitaxial film, it basically arises from the
strain, the composition, and the microstructure of the thin film. In the
present case, since all the samples were grown under identical conditions,
the increase in FWHM is mostly arising from the strains in the film. Thus,
it reveals that first, the strain induced in these heterostructures are
dependent on the BTO thickness layer and second, that it is maximum in the
case of the (5/15) superlattice (5 u.c. LCMO / 15 u.c. BTO).\ Moreover, the
decrease of the FWHM values at higher thickness of BTO\ suggests that there
is a relaxation of the strains in the films.

To investigate further the structure of the films, we have calculated the
in-plane (a) and out-of-plane (c) lattice parameters of the films from XRD
data by measuring the asymmetric reflections, namely $10l$ ( where $l$=1, 2,
3). The lattice parameters of the films plotted as a function of the BTO
layer thickness are shown in Fig 3b. The experimental error is about{\bf \ }$%
\pm $0.01A. Fig. 3b reveals that the out-of-plane lattice parameter is
approaching to the bulk value of the BTO lattice parameter (close to 4.1A)
as the number of BTO u.c. increases. However, the in-plane lattice parameter
is almost independent of the number of BTO u.c and does not show any clear
trend. Therefore, we have also plotted in Fig 3c, the ($c/a$) ratio of the
in-plane ($a$) and the out-of-plane ($c$) lattice parameters as a function
of number of BTO u.c. since it provides information about the distortion of
the perovskite structure. As shown in Fig 3c, the ($c/a$) ratio is varying
in the range of 0.96 to 1.04 depending on the number of BTO u.c. The error
(based on the errors of the lattice parameters) is close to $\pm $0.02A\ and
the ($c/a$) ratio is increasing from 0.97 to 1.04 when increasing the number
of BTO u.c. from 5 to 15, leading to a distortion larger than 1 above 5 u.c.
This suggests that the thickness of the individual BTO layer is playing a
very important role in governing the superlattice structures.

We are now trying to correlated these results with the properties. The
ferroelectricity in BTO depends on the details of TiO$_6$ polyhedra. In
other words, tetragonal-BTO is ferroelectric whereas it is paraelectric in
the cubic form. From the lattice parameter point of view, the lattice
parameters difference between cubic and tetragonal forms of BTO is not very
large. However, further investigation of the distortion of the TiO$_6$
octahedra-as inferred by the ($c/a$) ratio-yields the reason for the two
observed electrical forms. In the tetragonal BTO (c/a = 1.01) Ti is
undersized for the TiO$_6$ octahedra and will align ferroelectrically in the 
$c$-direction. On the other hand, in the cubic form of BTO, Ti is not
constrained to align only in a particular direction and a paraelectric state
is observed. Thus, the ferroelectricity in BTO depends on the degree of
distortion in TiO$_6$ octahedra To create the multiferroic superlattices,
one expects that the BTO should be ferroelectric in nature and it is
possible only if the total superlattice structure is showing a distorted
perovskite structure {\it i.e.} 'c/a' ratio should not be equal to 1. In
addition, the electrical polarization in thin films of BTO depends on their
thickness.\cite{Junqure} However, the superlattices composed of BTO and
oxide dielectrics have shown that even one unit cell of BTO possess the
electrical polarization.\cite{HNLee} Thus, the presence of the
ferroelectricity in superlattices will be dependent on the details of the
layered materials, their thickness, and the strain in the films. On the
contrary, the magnetic properties of the manganites are governed by the Mn$%
^{3+}$-O-Mn$^{4+}$ structure.\cite{hwang} In other words, by varying the
Mn-O bond lengths and Mn$^{3+}$-O-Mn$^{4+}$ bond angle, its possible to
control (or suppress) the ferromagnetism in manganites confirming that the
strain in the film plays a major role in determining the Mn$^{3+}$-O-Mn$^{4+}
$ structure. Consequently, for designing a superlattice for multiferroics,
it will be important to find an optimum stress/strain in the film, so that
both the ferroelectricity and ferromagnetism are present. Furthermore, the
ferromagnetic layer should be insulating in nature, otherwise due to the
high leakage current the ferroelectricity will be suppressed in the
structure. Thus, one has to find a delicate balance between the
strain/stress in the superlattices together with the properties of the
individual components. In the present case, the distortion of the
superlattices depends on the layer thickness of the BTO. As mentioned above
the distortion in (5/5) is in-plane whereas it is out of plane in the (5/15)
superlattice case. Keeping in mind the easy axis direction of BTO
polarization, the (5/15) superlattice should exhibit the maximum coupling.
Here, it is also worth noting that the easy axis of electrical polarization
of bulk BTO is parallel to the $c$-axis ({\it i.e.} the out-of-plane
direction). Thus, to have a large ferroelectricity, and consequently a
maximum multiferroic coupling in the case of LCMO/BTO, the sequence should
be (5/15). Finally, such results show that the lattice parameters and the ($%
c/a$) ratio (Fig 3b and 3c) extracted from the XRD data corroborate our
FWHM-rocking curve findings.

The investigation of the nature of strains and their distribution in the
samples were carried out by the conventional sin$^2\psi $ method, where $%
\psi $ is the angle between the lattice plane normal and the sample surface.
The sin$^2\psi $ method is non-destructive and commonly used technique to
investigate the elastic properties of the polycrystalline materials.\cite
{ICNoyan,Cullity} This techniques have been extended sucessfully to study
the average elsatic properties of the multilayer films under certain
approximation.\cite{padhan,schweitz,bocquet} It provides a good information
about the average strain distribution and their nature in the films. In the
bi-axial model for a cubic structure, the strain ($\varepsilon $) in the
film along the [$hkl$] can be defined as\cite{ICNoyan}

a = $%
{\displaystyle {d_{hkl}(\phi \psi )-d_o \overwithdelims() d_o}}%
=$ $\varepsilon _{11}-\varepsilon _{33}\sin ^2\psi +\varepsilon _{33}\
............(1)$

where $\phi $ is the angle between the projected lattice plane normal and
in-plane axis. The parameters d$_{hkl}$ and d$_o$ are the strained and
unstrained plane spacing of the samples, respectively. $\varepsilon _{11}$
and $\varepsilon _{33}$ are the in-plane and out-of plane strain component
of the film. To estimate the strain in-plane and, out-of-plane the $d_o$
value was calculated by assuming that all the samples have same Poisson's
ratio ($\nu $) and using the theoretical value (0.37) of manganites since,
the Poisson's ratio is in the range of 0.3-0.5 for perovskites.\cite{eck,rao}
However, we are well aware of the shortcomings of the present assumptions
that will be discussed hereafter. For investigating the strain in the film,
we have chosen a unique direction with constant $h$ and $k$ to measure the
diffracted X-ray intensity as well as $\psi $ from the reflection. Value of
the $\psi $ is sensitive to the alignment of the sample and to minimize it,
we have averaged over all $\phi $-directions. A typical d$_{10l}$ {\it vs.}
sin$^2\psi $ is shown in the inset of Fig 4a. It shows that d$_{10l}$ varies
linearly with respect to sin$^2\psi $ , which shows that the strains are
uniformly distributed in the film. The estimated d$_{103}$ of the
superlattices is plotted as a function of BTO u.c. and shown in Fig 4a. Fig
4a clearly shows that the d$_{10l}$ varies a little with the increase in BTO
u.c. The estimated value of the in-plane and out-of-plane strains as a
function of BTO u.c. are shown in Fig 4b. From Fig 4b, it is evident that
the film has large in-plane tensile strain and compressive strain in
out-of-plane direction. To understand Fig. 4b, we need to define the terms
superlattices, strained multilayers, and relaxed multilayer structures. In
theory, an ideal superlattice can be defined as a single extended crystal
having perfect registry with the orientation of the underneath layer and its
'$a$' and '$b$' lattice parameters are basically governed by the substrate
whereas the 'c' parameter is closed to the sum of the bulk '$c$' lattice
parameters of each stacked material employed for fabricating the
superlattice. However, in reality due to the lattice mismatch and the
cationic size of the various constituents generating various kinds of
defects, strain at the interfaces occured, and in practice a superlattice is
considered as strained multilayer structures{\it . }Thus, in case of
multilayer the nature of strain will be varying between the two types of
strains: the interface film/substrate and the interfaces of the distinct
layers used for creating them. Therefore, one may expect that variation of
strains in a multilayer system, whereas in the case of a fully relaxed
multilayer system, the lattice parameters of the film should be close to the
bulk materials (composed of identical compositions of the superlattices). In
other words, ideally a fully relaxed multilayer system will exhibit a
lattice parameter equivalent to their bulk counterpart. In the present case,
it is evident that the film has a large in-plane tensile stress and an
out-of-plane compressive strain and its origin can be understood as follows.
The lattice mismatch between LCMO/STO is -1.17\%, BTO-STO is + 2.2 \% and
LCMO-BTO is 3.8 \%, respectively. Thus, the nature of the strain will be
varying from interface to interface, because the substrate will induce a
compressive strain whereas the LCMO/BTO will induce either compressive or
tensile strain depending on the underneath layer. In addition, strains at
the interfaces of LCMO/BTO will be increasing due to the variation in
cationic sizes. Similar results have been obtained in the case of SrRuO$_3$%
/SrMnO$_3$\ superlattices.\cite{padhan} In these films, Padhan {\it et. al. }%
have demonstrated that the variation in cationic sizes of Ru$^{4+}$ and Mn$%
^{4+}$ are imposing strains in the perovskite lattice. Similarly to our
superlattices, they have shown that the values of $\varepsilon _{11}$
(in-plane stress) and $\varepsilon _{33}$\ (out-of-plane stress) depend on
the thickness details of the individual layers with the same order of
magnitude as LCMO/BTO\ superlattices. Thus, our results are comparable to
the other oxides-based superlattices. Furthermore, with increasing the BTO
thickness more than 15 u.c. the out-of-plane strain decreases, which will
result in the stiffening of the lattices that generates defects at the
interfaces. Hence our system will behave as a relaxed multilayer and a
similar trend has been shown by the FWHM plot (Fig.3a) and the out-of-plane
lattice variation (Fig.3b). Thus, the strain analysis corroborates our
lattice parameters/FWHM findings. In addition, the creation of defects can
result in rough interfaces, which might suppress the multiferroic coupling
in the films as illustrated later.

However, we are aware that there are some shortcomings in the present
technique for extracting the strains/stress of multilayer films. In this
methodology, one assumes a single layer structure and evaluates the {\it %
average} strain/stress in the film, because the XRD can not distinguish
between individual interfaces.\ As a result, the XRD patterns provide only
average information about all interfaces and layers. Consequently, it does
not give any information about strain/stress distribution at the individual
interfaces. However, in multilayer structures, it is evident that the strain
at the different interfaces will be different due to the lattice mismatch
between the substrate, and various type of oxides employed for growing the
superlattices. Despite of the above shortcoming, the present structural data
provide a good qualitative information about the strain distribution in the
film.

\subsection{Morphological studies}

As seen through the structural analysis, the strain and the lattice
parameters of the superlattices strongly depend on the BTO layer thickness.
The morphology were also examined using atomic force microscopy.
Morphological parameters, such as the root mean square (RMS) roughness of
the superlattices, were extracted from the 2 x 2 $\mu $m$^2$ AFM micrographs
and were found to be in the range of 0.3-0.6 nm. Note that the RMS values
are very close to one unit cell of superlattices indicating that the films
have smooth surfaces. Fig. 5 shows the 2D-AFM micrographs of two samples,
namely (5/5) and (5/15) superlattices . Fig 5 clearly reveals that the
samples are mostly free-from particulates. The particulate density was found
to be around 0.3/$\mu $m$^2.$ Furthermore, the AFM micrographs of the
superlattices show that the mounds are spherical in nature and they are very
uniform in size. Moreover, the average size of the mounds are found to be
dependent on the details of the superlattices. For example, in a (5/15)
superlattice, the average mound size is of the order of 65 nm whereas in a
(5/5) superlattice it is close 40 nm. The mound size in the films usually
depends on various factors, such as the deposition temperature, substrate, 
{\it etc}. However, in the present case all samples were grown under
identical conditions. Thus, the morphological variation in the superlattices
is arising basically from their structural variations.

\subsection{Physical properties}

The variations in the structure and morphology of the superlattices were
clearly evident in their physical properties. Detail physical properties of
these films and the experimental details have been reported elsewhere\cite
{mpsingh}. For the sake of discussion, we are plotting the magnetic and
magneto-electric properties of these superlattices as a function of number
of BTO unit cells employed for designing them and the result is shown in
Fig.6. It shows that the magnetization and magnetoresistance of the
superlattices increased with the progressive increase in the number of BTO
unit cells and attains a maximum in the case of (5/15) superlattices.
Moreover, it is worth to keep in mind that the magnetic layer in the sample
is LCMO. Therefore, the present enhancement in magnetization and
magnetoresistance of the superlattices may be explained, based on the
possible multiferroic coupling in these structures. This also consistent
with the structural point of view,\ i.e., the (5/15) superlattice possesses
the lattice distortion in the c-direction (Fig.3). With further increase in
BTO layer thickness above 15 u.c., magnetization suppressed, which can be
understood based on lattice stiffening that results in the relaxation of the
film structure and in turn create defects at the interfaces. Hence, the
presence of interfacial defects will suppress the multiferroic coupling in
the superlattices and results in the decrease of magnetizations and
magnetoresistance. Hence, the structure of the superlattices is playing an
important role in the determination of their multiferroic properties.

\section{Conclusion}

To summarize, we have successfully grown (BaTiO$_3$/La$_{0.7}$Ca$_{0.3}$MnO$%
_3$) superlattices on (001)-oriented SrTiO$_3$ by pulsed laser deposition
process. Despite the lattice mismatch between substrate and La$_{0.7}$Ca$%
_{0.3}$MnO$_3$ ($-1.17\%$), and BaTiO$_3$ ($+2.2\%$), the films were grown
heteroepitaxial. The structural analysis showed that the strains are
uniformly distributed in the samples and vary with the BaTiO$_3$ layer
thickness. In addition, it appears that first the strain induced in these
heterostructures are dependent on the BaTiO$_3$ thickness layer and second,
that it is maximum in the case of the (5/15) superlattice (5 u.c. LCMO / 15
u.c. BTO). The physical properties measurements indicated that there is a
multiferroic coupling in these structures and that the strains play a
particular role in optimizing such behavior. This study may provide a way to
design new multiferroics.

We would like to thank Prof. B.\ Mercey, Dr. P. Padhan, and Dr. H. Eng for
the helpful discussions.

This work has been carried out in the frame of the Work Package ''New
Architectures for Passive Electronics'' of the European Network of
Excellence ''Functionalized Advanced Materials Engineering of Hybrids and
Ceramics'' FAME (FP6-500159-1) supported by the European Community, and by
Centre National de la Recherche Scientifique.

Figure 1: (a) $\Theta $-2$\Theta $ XRD pattern of the superlattices with
varying BTO u.c. and (b) chemical modulation ($\Lambda )$ of the
superlattices as a function of BTO u.c. Solid line in Fig 1b is just a
visual guide.

Figure 2: (a) $\Theta $-2$\Theta $ XRD pattern and simulated XRD pattern of
the (5/10) superlattice (b) pole-figure of 5/10 superlattice recorded around
the $103$-reflection.

Figure 3: (a) FWHM of 002-rocking curve as function of BTO u.c., (b) '$a$'
and '$c$' lattice parameters of the films plotted as a function of BTO u.c.,
and (c) the variation of '$c/a$' ratio as a function of BTO u.c. Inset of
Fig 3a shows the rocking curve ($\omega $-scan) for ($5/10$) superlattice.
Solid lines are just visual guides.

Figure 4: (a) Variation of d$_{103}$ as a function of BTO u.c. and (b)
in-plane and out-of-plane stress as a function of the BTO u.c. Inset of Fig.
4a shows the sin$^2\psi $ {\it vs.} d$_{10l}$ plot for (5/10) superlattices
and the solid line is the linear fit to experimental data.

Figure 5: 2D-AFM micrographs of (a): (5/5) and (b): (5/15) superlattices.

Figure 6: Magnetic moment (solid circle) measured at 10 K and
magnetoresistance (open circle) measured at 100 K of films plotted as a
function of number of BTO u.c. The solid straight and dash line are just
visual guide.

\end{document}